\documentclass[conference,citesort]{IEEEtran}
\IEEEoverridecommandlockouts

\usepackage{cite,citesort,amsbsy,amsthm}
\ifCLASSINFOpdf
  \usepackage[pdftex]{graphicx}
  \graphicspath{{../pdf/}{../jpeg/}}
  \DeclareGraphicsExtensions{.pdf,.jpeg,.png}
\else
  \usepackage[dvipdfmx]{graphicx}
  \graphicspath{{../eps/}}
  \DeclareGraphicsExtensions{.eps}
\fi
\usepackage{amssymb,array,mdwmath,mdwtab}
\usepackage[cmex10]{amsmath}
\usepackage{enumerate,subfigure}
\usepackage{multicol,multirow,paralist}
\usepackage{color}
\usepackage{algorithm}
\usepackage{algorithmic}

\makeatletter
\newcommand{\removelatexerror}{\let\@latex@error\@gobble}
\makeatother
\begin{document}

\title{Polar Coding with Chemical Reaction Networks}

\author{
  \IEEEauthorblockN{
    Toshiki Matsumine\IEEEauthorrefmark{3}\IEEEauthorrefmark{2}\thanks{T. Matsumine conducted this research when he was an intern at MERL.},
    Toshiaki Koike-Akino\IEEEauthorrefmark{3},
    and Ye Wang\IEEEauthorrefmark{3}
  }
  \IEEEauthorblockA{
    \IEEEauthorrefmark{3}
    Mitsubishi Electric Research Laboratories (MERL), 
    201 Broadway, Cambridge, MA 02139, USA. \\
    \IEEEauthorrefmark{2}
    Department of Electrical and Computer Engineering,
    Yokohama National University, \\
    79-5 Tokiwadai, Hodogaya, Yokohama, Kanagawa, Japan\\
    Email: matsumine-toshiki-tk@ynu.jp,
    \{koike, yewang\}@merl.com
  }
}

\maketitle

\begin{abstract}
  In this paper, we propose a new polar coding scheme with molecular programming, which is capable of highly parallel implementation at a nano-scale without a need of electrical power sources. We designed chemical reaction networks (CRN) to employ either successive cancellation (SC) or maximum-likelihood (ML) decoding schemes for short polar codes. From ordinary differential equation (ODE) analysis of the proposed CRNs, we demonstrate that SC and ML decoding achieve accurate  computations across fully-parallel chemical reactions. We also make a comparison in terms of the number of required chemical reactions and species, where the superiority of ML decoder over SC decoder is observed for very short block lengths.

\end{abstract}

%
\IEEEpeerreviewmaketitle

\section{Introduction}
Thanks to recent advancements of computer science and bio-molecular technologies, molecular programming has been rapidly grown as an emerging topic. Chemical reaction networks (CRNs) are a useful descriptive programming language for modeling complex chemical systems. New possible applications with CRNs have been widely studied so far.
For example,  chemical implementation of neural networks has been studied in \cite{hjelmfelt1991chemical,kim2005neural,qian2011neural}. Also digital logic \cite{shea2010writing,jiang2011synchronous,jiang2013digital,senum2011rate,cardelli2018chemical} and belief propagation (BP) with CRNs have been proposed in \cite{napp2013message,xingchi2018synthesizing}. CRNs are a Turing-universal model, with which we can perform arbitrary computation \cite{soloveichik2008computation}. Although it is beyond our scope in this paper, deoxyribonucleic acid (DNA) has been extensively studied to translate CRNs in the literature \cite{shin2012compiling}. In fact, designing DNA strands can realize the entire dynamic behaviors of CRNs \cite{soloveichik2010dna}.

The proposed polar coding with chemical programming may be useful for intra-body communications, for human health monitoring and drug delivery \cite{tadashi2012molecular}. In such applications, the transmitted data size is expected to be small. Therefore, short channel coding that has powerful error correcting capability may be required for such applications.
As already shown in the literature \cite{liva2016code,mahyar2018short}, polar codes concatenated with cyclic redundancy check (CRC) have an advantage over low-density parity-check (LDPC) codes in short block-length regime, which indicates that polar codes may be a potential candidate for error-correcting codes in molecular communication systems.

In this paper, we propose several chemical implementations of short polar codes. We investigate two decoding schemes of short polar codes: successive cancellation (SC) decoding and maximum-likelihood (ML) decoding. Our implementation does not rely on the specific rate constant of chemical reactions, and thus parallel implementation is possible. From simulation results, we show that performances of two decoding schemes in terms of the trade-off between computation accuracy and speed are comparable, whereas ML decoder can be realized by using much smaller number of chemical species and reactions when the code length is very short.
The main contributions of this paper are summarized as follows:
\begin{itemize}
\item We propose a novel application of molecular computing to polar encoding and decoding;
\item We design two CRN-based decoding schemes of polar codes: SC and ML decoding; and
\item We evaluate the performance trade-off between accuracy and speed of the proposed CRNs, as well as the number of required chemical species and reactions.
\end{itemize}


\section{Preliminaries}
\label{sec:pre}

\subsection{Chemical Reaction Networks (CRNs)}
Consider an example of CRNs involving three chemical species $A$, $B$, and $C$, which follow a chemical reaction as below:
\begin{align}
  A + B \overset{k}{\rightarrow} C,
\end{align}
where $A$ and $B$ are {\em{reactants}}, $C$ is {\em{products}}, and $k$ is the rate constant that indicates how fast the reaction occurs. Species that participate in a reaction, but no consumption or production occurs are called {\em{catalysts}}. For example, the species $A$ in the following reaction is called catalyst 
\begin{align}
  A + B \overset{k}{\rightarrow} A + C.
\end{align}
In CRNs, empty set $\phi$ may be used as both reactants and products. In the following example, it is used as reactants
\begin{align}
  \phi \overset{k}{\rightarrow} C,
\end{align}
where $\phi$ means the products are generated from a large or replenishable source. When $\phi$ appears as products as follows:
\begin{align}
  A + B \overset{k}{\rightarrow} \phi,
\end{align}
this reaction means that species $A$ and $B$ cancel out equal concentrations by transferring them to an external sink.
We design CRNs so that the {\em{equilibrium}} concentration of some species shows the result that we want to compute.
Throughout this paper, we assume the reaction rate of $1$, since our implementation does not depend on the specific rate.

\subsection{Basic Operations}
In order for the chemical implementation of polar encoding and decoding, both bit and probability computation with CRNs are required. In this subsection, we review how bits or probabilities are represented using a molecular concentration and some basic computations with CRNs\cite{xingchi2018synthesizing} that we use in this work.

\subsubsection{Bit Representation}
In order to represent a bit with CRNs, we use a {\em{complementary representation}} \cite{jiang2013digital}, where two molecular $A^0$ and $A^1$ are used for a single bit $A$. The presence of molecular $A^0$ indicates that $A=0$, and vice versa. For this reason, molecular $A^0$ and $A^1$ should not exist at the same time. This is implemented by the following reaction set
\begin{align}
  A^0 + A^1 &\rightarrow S, \label{eq:bit} \\
  S + A^0 &\rightarrow 3 A^0, \\
  S + A^1 &\rightarrow 3 A^1,
\end{align}
where in \eqref{eq:bit}, $A^0$ and $A^1$ are consumed so they exist exclusively in steady states.

\subsubsection{Probability Expression}
The probability is expressed by a ratio of concentrations of two molecules $A^0$ and $A^1$,
\begin{align}
  P_A &= \frac{[A^1]}{[A^0]+[A^1]}, \\
  P^\mathrm{c}_A &= 1-P_A = \frac{[A^0]}{[A^0]+[A^1]},
\end{align}
where $[\cdot]$ denotes the concentration of the argument molecule.

\subsubsection{Probability Multiplication}
Let us consider multiplication of two probabilities $P_C = P_A \times P_B$. Letting initial concentrations $[C^0]=[C^1]=0.5$, we transfer $[C^0]$ to $[C^1]$ when $AB=1$ and $[C^1]$ to $[C^0]$ when $AB=0$. This is performed by the following set of reactions
\begin{align}
  A^0 + C^1 &\rightarrow A^0 + C^0, \label{eq:mul1} \\
  A^1 + B^0 + C^1 &\rightarrow A^1 + B^0 + C^0, \label{eq:mul2} \\
  A^1 + B^1 + C^0 &\rightarrow A^1 + B^1 + C^1, \label{eq:mul3}
\end{align}
where in \eqref{eq:mul1}, $[C^1]$ is transformed to $[C^0]$ for $A=0$. Similarly, \eqref{eq:mul2} corresponds to $A=1, B=0$, and \eqref{eq:mul3} corresponds to $A=1, B=1$.

In what follows, we denote this probability multiplication by
\begin{align}
  C^0, C^1 = \mathsf{multiply}(A^0, A^1, B^0, B^1).
\end{align}
Our implementation is different from that in \cite{xingchi2018synthesizing} in that we continuously calculate $P_C = P_A \times P_B$, whereas the calculation in \cite{xingchi2018synthesizing} is performed only when an auxiliary molecule $S$ exists.

\subsubsection{Probability Division}
Suppose division of two probabilities $P_C = P_A / (P_A + P_B)$, we initialize concentrations of two molecules as $[C^0]=[C^1]=0.5$. Then division is performed by transferring $[C^0]$ and $[C^1]$ such that their ratio is equal to that of $[A^1]$ and $[B^1]$ , i.e., $[A^1]:[B^1]=[C^0]:[C^1]$. The following set of reactions performs division of two probabilities, $P_C = [C^1]/([C^0]+[C^1]) = P_A / (P_A + P_B)$:
\begin{align}
  A^1 + C^1 &\rightarrow A^1 + C^0, \\
  B^1 + C^0 &\rightarrow B^1 + C^1.
  \label{eq:mul}
\end{align}
For simplicity, we use the following notation for this function
\begin{align}
  C^0, C^1 = \mathsf{divide}(A^0, A^1, B^0, B^1).
\end{align}

\section{Chemical Polar Encoder}
\label{sec:enc}
In this section, our proposed chemical implementation of polar encoding is described. The generator matrix of polar codes with a block length of $N$ is expressed as follows:
\begin{align}
  G = \left[
  \begin{array}{cccc}
    1 & 0 \\
    1 & 1 
  \end{array}
        \right] ^ {\otimes n},
        \label{eq:enc}
\end{align}
where $[\cdot]^{\otimes n}$ denotes the $n$th Kronecker power with $n = \log_2 N$.

As we can see from \eqref{eq:enc}, the polar encoder is constructed from basic two bit-wise operations, which are copy and exclusive-OR (XOR). Suppose $A$ and $B$ are input bits, a chemical implementation of XOR operation $C = \mathsf{XOR}(A, B)$ is proposed in \cite{jiang2013digital} as below:
\begin{align}
  A^0 + B^1 &\rightarrow A^0 + B^1 + {C^1}', \label{eq:xor1} \\
  A^1 + B^0 &\rightarrow A^1 + B^0 + {C^1}', \\
  {C^1}' &\rightarrow \phi, \\
  {C^1}' + C^0 &\rightarrow C^1, \label{eq:xor2} \\
  A^0 + B^0 &\rightarrow A^0 + B^0 + {C^0}', \label{eq:xor3} \\
  A^1 + B^1 &\rightarrow A^1 + B^1 + {C^0}', \\
  {C^0}' &\rightarrow \phi, \\
  {C^0}' + C^1 &\rightarrow C^0, \label{eq:xor4}
\end{align}
where reactions in \eqref{eq:xor1}--\eqref{eq:xor2} will generate the molecular $C^1$ corresponding to $A=0, B=1$ and $A=1, B=0$, and reactions \eqref{eq:xor3}--\eqref{eq:xor4} correspond to $A=0, B=0$ and $A=1, B=1$. Copy operation can be performed by reusing XOR implementation. Specifically, we perform copy of $A$ by taking XOR with "0", i.e.,  $C=\mathsf{copy}(A)=\mathsf{XOR}(A, 0)$.

\section{Chemical SC Decoding}
\label{sec:sc}
In this section, we describe the proposed implementation of polar decoder based on factor graph. Fig.~\ref{fig:sc} shows the factor graph of polar codes with a block length of $N=4$. As shown in this figure, SC decoding consists of two fundamental computations, which are f-function and g-function \cite{tehrani2006stochastic}. In what follows, we review these functions to propose efficient implementations with CRNs. Although f- and g-functions for LDPC BP decoders are proposed in \cite{xingchi2018synthesizing}, our approach is different from \cite{xingchi2018synthesizing} in that our polar decoder is asynchronous and does not require chemical clock CRNs for circuit synchronization.
\begin{figure}[t]
  \centering
  \includegraphics[width=0.9\hsize]{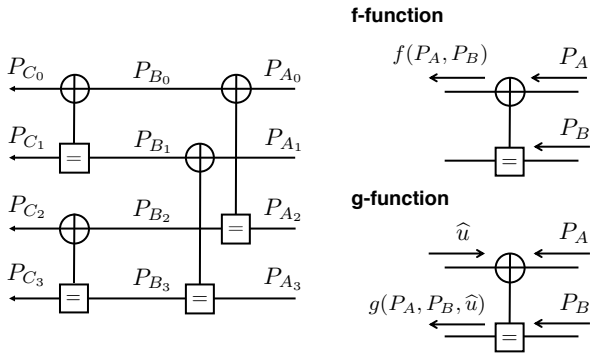}
  \caption{Polar decoder with block length $N=4$. Probability is represented by ratio of concentrations of two molecules $P_{A_i}=[A^1_i]/([A^1_i]+[A_i^0])$.}
  \label{fig:sc}
\end{figure}

\subsection{The f-Function}
The f-function in SC decoding takes two probabilities $P_A$ and $P_B$ as its input, and outputs $P_C=f(P_A, P_B)$. As shown in \cite{tehrani2006stochastic}, the probability $P_C$ and its complementary probability $P^\mathrm{c}_C = 1-P_C$ is calculated as 
\begin{align}
  P_C &= P_A (1 - P_B) + P_B (1 - P_A), \label{eq:f1} \\ 
  P^\mathrm{c}_C &= 1-P_C = (1 - P_A) (1 - P_B) + P_A P_B. \label{eq:f2}
\end{align}
Initializing molecular concentrations as $[C^0]=[C^1]=0.5$, the f-function is implemented by the following set of reactions:
\begin{align}
  D =& \mathsf{multiply}(P_A, P_B), \label{eq:f3} \\
  E =& \mathsf{multiply}(P^\mathrm{c}_A, P_B), \label{eq:f4} \\
  F =& \mathsf{multiply}(P_A, P^\mathrm{c}_B), \label{eq:f5} \\
  G =& \mathsf{multiply}(P^\mathrm{c}_B, P^\mathrm{c}_B), \label{eq:f6} \\
  F + C^0 &\rightarrow F + C^1, \label{eq:f7} \\
  E + C^0 &\rightarrow E + C^1, \label{eq:f8} \\
  D + C^1 &\rightarrow D + C^0, \label{eq:f9} \\
  G + C^1 &\rightarrow G + C^0, \label{eq:f10}
\end{align}
where $P_D$, $P_E$, $P_F$, and $P_G$ correspond to $P_A P_B$, $P^\mathrm{c}_A P_B$, $P_A P^\mathrm{c}_B$, and $P^\mathrm{c}_A P^\mathrm{c}_B$, respectively.
After that, reactions \eqref{eq:f7}--\eqref{eq:f10} transfer $C^0$ to $C^1$ and vice versa, and the resulting concentrations can realize the computations \eqref{eq:f1} and \eqref{eq:f2}.

\subsection{The g-Function}
In addition to two input probabilities $P_A$ and $P_B$, the g-function depends on the decision $\widehat{u}$ after the f-function. We denote g-function by $P_C=g(P_A, P_B, \widehat{u})$, which is calculated as follows:
\begin{align}
  P_C =
  \begin{cases}
    \frac{P_A P_B}{P_A P_B + (1 - P_A)(1 - P_B)},   & \text{if} \ \widehat{u}=0, \\
    \frac{(1 - P_A) P_B}{(1 - P_A) P_B + P_A(1 - P_B)},  & \text{otherwise}.
  \end{cases}
\end{align}

The following set of reactions shows the computation of the g-function when $\widehat{u}=0$:
\begin{align}
  P_D &= \mathsf{multiply}(P_A, P_B), \label{eq:g1} \\
  P_G &= \mathsf{multiply}(P^\mathrm{c}_A, P^\mathrm{c}_B), \label{eq:g2} \\
  P_C &= \mathsf{divide}(P_D, P_G). \label{eq:g3} 
\end{align}
Analogously, the g-function with $\widehat{u}=1$ can be performed by replacing $P_A$ and $P^\mathrm{c}_A$ in reactions \eqref{eq:g1} and \eqref{eq:g2}.

Letting $A^0$ and $A^1$ denote outputs from the f-function, we make a decision and set $[U^0]=1, [U^1]=0$ if $[A^0]>[A^1]$ and otherwise $[U^0]=0, [U^1]=1$. This is implemented by following reactions:
\begin{align}
  A &\rightarrow A + X, \label{eq:g4} \\
  A^\mathrm{c} &\rightarrow A^\mathrm{c} + Y, \label{eq:g5} \\
  X + Y &\rightarrow \phi, \label{eq:g6} \\
  X + U^0 &\rightarrow X + U^1, \label{eq:g7} \\
  Y + U^1 &\rightarrow Y + U^0, \label{eq:g8} 
\end{align}
where initial concentrations of $U^0$ and $U^1$ are $[U^0]=[U^1]=0.5$. Reactions \eqref{eq:g4} and \eqref{eq:g5} copy $A^0$ and $A^1$ to $X$ and $Y$, respectively, and \eqref{eq:g7} and \eqref{eq:g8} divides $U^0$ and $U^1$ according to the ratio $[A^0]:[A^1]$. Reaction \eqref{eq:g6} consumes molecules $X$ and $Y$, such that either of them is completely consumed. The reactions \eqref{eq:g4}--\eqref{eq:g8} are not required for frozen bits, where we set $[U^0]=1, [U^1]=0$. 

As mentioned earlier, since the g-function depends on the previous decision, we implement this as follows:
\begin{align}
  U^0 + g(P_A, P_B, \widehat{u}=0) \rightarrow U^0 + g(P_A, P_B, \widehat{u}=0), \label{eq:g9} \\
  U^1 + g(P_A, P_B, \widehat{u}=1) \rightarrow U^1 + g(P_A, P_B, \widehat{u}=1). \label{eq:g10}
\end{align}
Since either $U^0$ or $U^1$ is completely consumed in \eqref{eq:bit}, either of \eqref{eq:g9} and \eqref{eq:g10} is computed, e.g., only the reaction \eqref{eq:g9} occurs when $[U^0] > [U^1]$ \eqref{eq:g10} occurs otherwise.

\subsection{Parallel Asynchronous SC Decoding}
For traditional SC decoding of polar codes, parallel implementation is difficult, since the f-function should be calculated before the g-function since the g-function depends on the decision result from the f-function. On the other hand, our chemical SC decoder performs asynchronous decoding operations fully in parallel, i.e., all the f and g-functions are continuously calculated.

To do this, we continuously perform bit decision and polar encoding at the same time as decoding. This enables that the g-function is automatically updated based on the intermediate decision result of the f-function, and eventually decoding results are converged to the target results same as the conventional synchronous-circuit SC decoding.

\section{Chemical ML Decoding}
\label{sec:ml}
In this section, we describe the CRN design for ML decoding. The objective of ML decoding is to maximize the following probability
\begin{align}
  p(\mathbf{y}|\mathbf{x}) = \prod_i p(y_i|x_i),
  \label{eq:sym-like}
\end{align}
where $x_i$ and $y_i$ are the $i$th transmitted and received codeword bits, respectively. While optimal ML decoder finds the most-likely codeword that maximizes \eqref{eq:sym-like}, since the efficient implementation of max function of multiple variables with CRNs is difficult. Instead, we propose the sub-optimal ML decoding that maximizes the bit-wise likelihood, rather than symbol-wise likelihood in this paper.

The probability that the $i$th bit is $b$ is calculated as follows:
\begin{align}
  P^b_i = \frac{\sum_{\mathbf{x} \in \mathcal{X}^b_i} p(\mathbf{y}|\mathbf{x})}{\sum_{\mathbf{x} \in \mathcal{X}^0_i} p(\mathbf{y}|\mathbf{x}) + \sum_{\mathbf{x} \in \mathcal{X}^1_i} p(\mathbf{y}|\mathbf{x})},
  \label{eq:bit-like}
\end{align}
where $\mathcal{X}^b_i$ is a set of codewords whose $i$th element is $b \in \{0, 1\}$. In what follows, we describe how to calculate \eqref{eq:bit-like} with CRNs.

Here we take (4, 2) polar codes as an example for simplicity of explanations, whose generator matrix is given by 
\begin{align}
  G = \left[
  \begin{array}{cccc}
    0 & 0 & 1 & 1 \\
    1 & 1 & 1 & 1
  \end{array}
                \label{eq:gen}
                \right],
\end{align}
whose possible codewords are $\{0000, 0011, 1100, 1111\}$. We first consider computing \eqref{eq:sym-like}. In order to compute multiplication of multiple variables (more than two), 
we recursively perform multiplication of two variables. More specifically, multiplication of $N$ variables consists of $\log_2 N$ multiplications of two variables, i.e., $\mathsf{multiply}(P_{A_0}, P_{A_1}, P_{A_2}, P_{A_3})=\mathsf{multiply}(P_{A_0}, \mathsf{multiply}(P_{A_1}, \mathsf{multiply}(P_{A_2}, P_{A_3})))$. In this way, \eqref{eq:sym-like} is implemented as follows
\begin{align}
  B &= \mathsf{multiply}(P^\mathrm{c}_{A_0}, P^\mathrm{c}_{A_1}, P^\mathrm{c}_{A_2}, P^\mathrm{c}_{A_3}), \\
  C &= \mathsf{multiply}(P^\mathrm{c}_{A_0}, P^\mathrm{c}_{A_1}, P_{A_2}, P_{A_3}), \\
  D &= \mathsf{multiply}(P_{A_0}, P_{A_1}, P^\mathrm{c}_{A_2}, P^\mathrm{c}_{A_3}), \\
  E &= \mathsf{multiply}(P_{A_0}, P_{A_1}, P_{A_2}, P_{A_3}),
\end{align}
where molecular concentrations $[B], [C], [D], [E]$ correspond to the probability of each codeword, $0000, 0011, 1100, 1111$, respectively.

Finally, let $L^b_i$ correspond to the probability that $i$th data bit is $b$. Setting initial concentrations $[L^0_0]=[L^1_0]=[L^0_1]=[L^1_1]=0.5$, the bit-wise likelihood in \eqref{eq:bit-like} is calculated by the following reactions,
\begin{align}
  B + L^1_0 &\rightarrow B + L^0_0, \label{eq:bit1} \\
  C + L^1_0 &\rightarrow C + L^0_0, \label{eq:bit2} \\
  D + L^0_0 &\rightarrow D + L^1_0, \label{eq:bit3} \\
  E + L^0_0 &\rightarrow E + L^1_0, \label{eq:bit4} \\
  B + L^1_1 &\rightarrow B + L^0_1, \label{eq:bit5} \\
  C + L^0_1 &\rightarrow C + L^1_1, \label{eq:bit6} \\
  D + L^1_1 &\rightarrow D + L^0_1, \label{eq:bit7} \\
  E + L^0_1 &\rightarrow E + L^1_1, \label{eq:bit8} 
\end{align}
where reactions \eqref{eq:bit1}--\eqref{eq:bit4} correspond to the first data bit and those in \eqref{eq:bit5}--\eqref{eq:bit8} correspond to the second data bit. More specifically, a set of reactions \eqref{eq:bit1}--\eqref{eq:bit4} calculate $P^0_0 = (P_B + P_C)/(P_B + P_C + P_D + P_E)$, and \eqref{eq:bit1}--\eqref{eq:bit4} calculate $P^0_1 = (P_B + P_D)/(P_B + P_C + P_D + P_E)$.

\section{Simulation Results}
\label{sec:sim}
In this section, we compare the performances of our chemical SC and ML decoders for short polar codes in terms of the number of required chemical species and reactions, and also the trade-off between computation accuracy and speed.
To do so, we analyze the time evolution of proposed chemical reactions governed by associated ordinary differential equations (ODE)\cite{chempy}.  
We consider decoding polar codes with generator matrix of \eqref{eq:gen}. This code can be derived from \eqref{eq:enc} by setting first and third rows to frozen indices.

\begin{figure}[t]
  \centering
  \includegraphics[width=0.8\hsize]{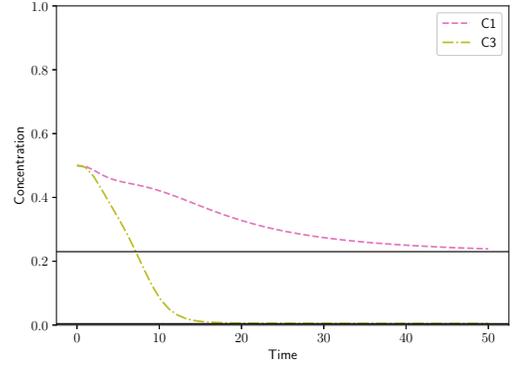}
  \caption{Convergence behavior of the molecular concentration over time with chemical SC decoder. Only the molecular concentration corresponding to the probability of $1$ is shown. Horizontal lines indicate the idealistic values.}
  \label{fig:sc}
\end{figure}
\begin{figure}[t]
  \centering
  \includegraphics[width=0.8\hsize]{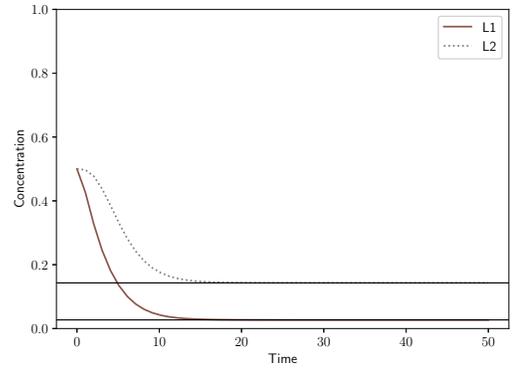}
  \caption{Convergence behavior of the molecular concentration over time with chemical ML decoder. Only the molecular concentration corresponding to the probability of $1$ is shown. Horizontal lines indicate the idealistic values.}
  \label{fig:ml}
\end{figure}

\subsection{Accuracy vs. Speed Trade-off}

Fig.~\ref{fig:sc} shows the result with the proposed SC decoder, where the probability vector $A_1=0.2, A_2=0.4, A_3=0.1, A_4=0.2$ is fed into the decoder. The expected decoder output in this case is $C_1=0.23, C^\mathrm{c}_1=0.77, C_3=0.0005, C^\mathrm{c}_3=0.9995$. From this figure, it is observed that molecular concentrations are converging to expected values, and hence it was demonstrated that our chemical polar decoder works well as intended.

Fig.~\ref{fig:ml} shows the result with the proposed chemical ML decoder. From \eqref{eq:sym-like} and \eqref{eq:bit-like}, the expected output vector is calculated as $L_0=0.143, L^\mathrm{c}_0=0.857, L_1=0.027, L^\mathrm{c}_1=0.973$. We can verify that the molecular concentrations converge to the expected values in Fig.~\ref{fig:ml}.

Table~\ref{tab1} summarizes the evolution of molecular concentration over time in SC and ML decoding. From this table, we observe that for both SC and ML decoding CRNs, molecular concentrations can converge to expected values as time proceeds. It is also observed that the convergence of molecule $C_2$ of SC decoding is slower than that of ML decoding, which stems from the delay associated with the decision of the previous bit used in the g-function.
\begin{table}[t]
  \centering
  \caption{Evolution of molecular concentration over time and comparison with expected values.}
  \begin{tabular}{c|c|c|c|c|c|c} \hline
    & Molecule & 0 [s] & 10 [s] & 20 [s] & 30 [s] & Expected \\ \hline
    \multirow{2}{*}{SC} & C1 & 0.500 & 0.420 & 0.325 & 0.272 & 0.249 \\ \cline{2-7}
    & C3 & 0.500 & 0.078 & 0.006 & 0.005 & 0.005 \\ \hline
    \multirow{2}{*}{ML} & L1 & 0.500 & 0.042 & 0.027 & 0.027 & 0.027 \\ \cline{2-7}
    & L2 & 0.500 & 0.174 & 0.143 & 0.143 & 0.143 \\ \hline
  \end{tabular}
  \label{tab1}
\end{table}

\subsection{Number  of Chemical Species and Reactions}
Finally, we briefly compare two decoding schemes in terms of the required number of chemical species and reactions. Table~\ref{tab2} shows the number of chemical reactions and species required in the proposed SC and ML decoding for half-rate polar codes with code lengths of $N=4, 8, 16$. From these results, we can see that ML decoder requires much less chemical reactions and species even with the relatively higher-speed convergence when the code length is very short $N=4$, whereas it increases exponentially as the code length increases. We conclude from this result that ML decoding is better when the code length is very small or faster processing speed is required, however, SC decoding may be suited when the code length is longer, e.g., greater than $N=8$ in terms of the CRN-based circuit size.
\begin{table}[t]
  \centering
  \caption{Number of chemical reactions and species required in the proposed CRN-based SC and ML decoding of half-rate polar codes with code length $N$.}
  \begin{tabular}{c|c|c} \hline
    & $\#$ of Reactions ($N=4, 8, 16$) & $\#$ of Species ($N=4, 8, 16$)\\ \hline
    SC & 222, 640, 1704 &  124, 356, 912 \\ \hline
    ML & 44, 224, 4352  &  36, 152, 2608 \\ \hline
  \end{tabular}
  \label{tab2}
\end{table}

\section{Conclusion}
\label{sec:con}
In this paper, we have proposed new polar coding scheme based on molecular programming. Our polar coding does not rely on clock, and thus enables highly parallel implementation. Two decoding schemes for short polar codes based on SC and ML decoding have been investigated. ODE simulation results demonstrated that our proposed CRNs of SC and ML decoders can realize fully-parallel and accurate decoding over chemical reactions without a need of power supply, while ML decoder requires much less chemical species and reactions when the code length is very short. To the best of our knowledge, our research is the first attempt to implement polar encoding and decoding via CRNs, which can be used for intra-body molecular communications without electric energy supplies.

We have focused on a specific short polar code and hence efficient generalization to longer codes should be studied.
Future work also includes the design of DNA strands that realize our chemical polar decoder. The {\em{cello}} \cite{nielsen2016genetic} may be useful for programming our chemical polar decoders using synthetic genetic circuits.


\ifCLASSOPTIONcaptionsoff
\newpage
\fi

\bibliographystyle{IEEEtran}
\bibliography{IEEEabrv,matsumine}

\begin{thebibliography}{10}
\providecommand{\url}[1]{#1}
\csname url@samestyle\endcsname
\providecommand{\newblock}{\relax}
\providecommand{\bibinfo}[2]{#2}
\providecommand{\BIBentrySTDinterwordspacing}{\spaceskip=0pt\relax}
\providecommand{\BIBentryALTinterwordstretchfactor}{4}
\providecommand{\BIBentryALTinterwordspacing}{\spaceskip=\fontdimen2\font plus
\BIBentryALTinterwordstretchfactor\fontdimen3\font minus
  \fontdimen4\font\relax}
\providecommand{\BIBforeignlanguage}[2]{{%
\expandafter\ifx\csname l@#1\endcsname\relax
\typeout{** WARNING: IEEEtran.bst: No hyphenation pattern has been}%
\typeout{** loaded for the language `#1'. Using the pattern for}%
\typeout{** the default language instead.}%
\else
\language=\csname l@#1\endcsname
\fi
#2}}
\providecommand{\BIBdecl}{\relax}
\BIBdecl

\bibitem{hjelmfelt1991chemical}
A.~Hjelmfelt, E.~D. Weinberger, and J.~Ross, ``Chemical implementation of
  neural networks and turing machines,'' \emph{Proceedings of the National
  Academy of Sciences}, vol.~88, no.~24, pp. 10\,983--10\,987, 1991.

\bibitem{kim2005neural}
J.~Kim, J.~Hopfield, and E.~Winfree, ``Neural network computation by in vitro
  transcriptional circuits,'' in \emph{Advances in neural information
  processing systems}, pp. 681--688, 2005.

\bibitem{qian2011neural}
L.~Qian, E.~Winfree, and J.~Bruck, ``Neural network computation with {DNA}
  strand displacement cascades,'' \emph{Nature}, vol. 475, no. 7356, p. 368,
  2011.

\bibitem{shea2010writing}
A.~Shea, B.~Fett, M.~D. Riedel, and K.~Parhi, ``Writing and compiling code into
  biochemistry,'' in \emph{Proc. the Pacific Symposium on Biocomputing}, pp.
  456--464, 2010.

\bibitem{jiang2011synchronous}
H.~Jiang, M.~Riedel, and K.~Parhi, ``Synchronous sequential computation with
  molecular reactions,'' in \emph{Proceedings of the 48th Design Automation
  Conference}, pp. 836--841.\hskip 1em plus 0.5em minus 0.4em\relax ACM, 2011.

\bibitem{jiang2013digital}
H.~Jiang, M.~D. Riedel, and K.~K. Parhi, ``Digital logic with molecular
  reactions,'' in \emph{Proceedings of the International Conference on
  Computer-Aided Design}, pp. 721--727, 2013.

\bibitem{senum2011rate}
P.~Senum and M.~Riedel, ``Rate-independent constructs for chemical
  computation,'' \emph{PloS one}, vol.~6, no.~6, 2011.

\bibitem{cardelli2018chemical}
L.~Cardelli, M.~Kwiatkowska, and M.~Whitby, ``Chemical reaction network designs
  for asynchronous logic circuits,'' \emph{Natural computing}, vol.~17, no.~1,
  pp. 109--130, 2018.

\bibitem{napp2013message}
N.~E. Napp and R.~P. Adams, ``Message passing inference with chemical reaction
  networks,'' in \emph{Advances in Neural Information Processing Systems}, pp.
  2247--2255, 2013.

\bibitem{xingchi2018synthesizing}
Z.~Xingchi, G.~Lulu, Y.~Xiaohu, and Z.~Chuan, ``Synthesizing {LDPC} belief
  propagation decoding with molecular reactions,'' in \emph{Proc. 2018 {IEEE}
  International Conference on Communications (ICC)}, Jul. 2018.

\bibitem{soloveichik2008computation}
D.~Soloveichik, M.~Cook, E.~Winfree, and J.~Bruck, ``Computation with finite
  stochastic chemical reaction networks,'' \emph{natural computing}, vol.~7,
  no.~4, pp. 615--633, 2008.

\bibitem{shin2012compiling}
S.~W. Shin, ``Compiling and verifying {DNA}-based chemical reaction network
  implementations,'' Ph.D. dissertation, California Institute of Technology,
  2012.

\bibitem{soloveichik2010dna}
D.~Soloveichik, G.~Seelig, and E.~Winfree, ``{DNA} as a universal substrate for
  chemical kinetics,'' \emph{Proceedings of the National Academy of Sciences},
  vol. 107, no.~12, pp. 5393--5398, 2010.

\bibitem{tadashi2012molecular}
T.~Nakano, M.~J. Moore, F.~Wei, A.~V. Vasilakos, and J.~Shuai, ``Molecular
  communication and networking: Opportunities and challenges,'' \emph{{IEEE}
  Trans. Nanobiosci.}, vol.~11, no.~2, pp. 135--148, 2012.

\bibitem{liva2016code}
G.~Liva, L.~Gaudio, T.~Ninacs, and T.~Jerkovits, ``Code design for short
  blocks: A survey,'' \emph{arXiv:1610.00873}, Oct. 2016.

\bibitem{mahyar2018short}
M.~Shirvanimoghaddam, M.~S. Mohamadi, R.~Abbas, A.~Minja, C.~Yue, B.~Matuz,
  G.~Han, Z.~Lin, Y.~Li, S.~Johnson, and B.~Vucetic, ``Short block-length codes
  for ultra-reliable low-latency communications,'' \emph{arXiv:1802.09166},
  Sep. 2018.

\bibitem{tehrani2006stochastic}
S.~S. Tehrani, W.~J. Gross, and S.~Mannor, ``Stochastic decoding of {LDPC}
  codes,'' \emph{{IEEE} Commun. Lett.}, vol.~10, no.~10, pp. 716--718, 2006.

\bibitem{chempy}
B.~Dahlgren, ``{ChemPy}: A package useful for chemistry written in {Python},''
  \emph{Journal of Open Source Software}, vol.~3, no.~24, p. 565, 2018.

\bibitem{nielsen2016genetic}
A.~A. Nielsen, B.~S. Der, J.~Shin, P.~Vaidyanathan, V.~Paralanov, E.~A.
  Strychalski, D.~Ross, D.~Densmore, and C.~A. Voigt, ``Genetic circuit design
  automation,'' \emph{Science}, vol. 352, no. 6281, p. aac7341, 2016.

\end{thebibliography}

\end{document}